\documentclass[aps,12pt,pra,showpacs,preprint,a4paper]{revtex4}
\usepackage{amsmath}
\usepackage{subfig}
\usepackage{array}
\usepackage{amsfonts}
\usepackage{epsf}
\usepackage{epsfig}
\usepackage{amssymb}
\usepackage{bbm}
\numberwithin{equation}{section}

\begin{document}
\title{Few quantum mechanical models in higher dimension-framed with Elzaki transform}
\author{\small Tapas Das}
\email[E-mail: ]{tapasd20@gmail.com}\affiliation{Kodalia Prasanna
Banga High School (H.S), South 24 Parganas, 700146, India}
\begin{abstract}
Very first time Elzaki transform is used in non relativistic quantum mechanics to solve $N$-dimensional Schr\"{o}dinger equation in a closed form for different solvable potential models. A universal transformation scheme is introduced and a formula based approach is developed which shows how to apply Elzaki transform to differential equation with non-constant coefficients that generally appear in solving quantum mechanical initial value problems specially for multidimensional Schr\"{o}dinger equation.   
\end{abstract}
 \maketitle
\section{I\lowercase{ntroduction}}
Symbolically integral equation appears as 
\begin{eqnarray*}
\eta(x)f(x)=g(x)+\lambda\int_a^b\kappa(x,y)f(y)dy\,,
\end{eqnarray*}
where $\eta(x), f(x), \kappa(x,y)$ are given and $f(x)$ is looking for a solution. The quantity $\lambda$ is a parameter which may be complex in general. The bi-variate function $\kappa(x,y)$ is called the kernel of the integral equation and it is defined and continuous on $a\leq x\leq b$ and  $a\leq y\leq b$. The other functions $\eta(x), f(x)$ are also defined and continuous on $a\leq x\leq b$. If $\eta(x)=0$ and $\lambda=-1$ we obtain the Fredholm equation [1-2] of first kind
\begin{eqnarray*}
g(x)=\int_a^b\kappa(x,y)f(y)dy\,.
\end{eqnarray*}  
This equation is popularly known as integral transform. Clearly, the possible types of integral transforms are unlimited depending on the choice of $\kappa(x,y)$. Laplace, Fourier, Mellin, Hankel transforms are famous and they are used in mathematical analysis and in physical application.\\
In this paper we will use a new transformation, called Elzaki transform, to analyze the non relativistic quantum mechanical problems. Compared to Fourier transform the Laplace transform has the advantage that it incorporates the initial condition in the transformation. That is the reason of extensive study of Schr\"{o}dinger equation, both for lower and multidimensional cases, via Laplace transformation [3-12]. Though, Laplace transform provides closed form solution but it has own disadvantage. Laplace transform of a differential equation with variable coefficient with power $i>1$ is not useful because the order of transformed differential equation is not reduced as well the transformed problem becomes much more difficult than in real space. This disadvantage motivated researchers to develop other integral transform like Sumudu [13-14] and Elzaki [15-16]. For example Euler-Cauchy equation [$x^2y^{''}(x)+axy^{'}(x)+by=0$ where $a,b$ are constant] is difficult to solve in Laplace transform but it is comparatively easy by Elzaki transform [17]. In this paper we will concentrate on the Elzaki transform and show that how it is effortlessly solve Schr\"{o}dinger equation for various potential models. Towards the motive we have generated a complete guideline, for applying this very recent transform, which will surely demand a position with the other analytical methods.\\
The paper is arranged as follows: The next section is for a brief of Elzaki transform where a general scheme of transformation technique for the solution of multidimensional Schr\"{o}dinger equation are illustrated step by step. Section 3 is devoted for the application of the method where we have retrieved some well known results for the spectrum of multidimensional Schr\"{o}dinger equation for few most celebrated potentials. Conclusion of the work comes at section 4. Important references are given after the conclusion. Finally we have added appendix where table of Elzaki transform of few functions is supplied and also few related important mathematics are furnished.
\section{A \lowercase{brief of} E\lowercase{lzaki transform}}
The Elzaki transform of the functions belonging to a class $\mathcal{A}$, where\\ $\mathcal{A}=\left\{f(t)|\exists M,q_1,q_2>0 \,\,such\,\ that\,\ |f(t)|<M e^{\frac{|t|}{q_j}}, \,\ if\,\ t\in (-1)^j\times[0,\infty)\right\}$\\
where $f(t)$ is denoted by $E[f(t)]=T(u)$ and defined as
\begin{eqnarray}
T(u)=u^2\int_0^{\infty}f(ut)e^{-t}dt=u\int_0^{\infty}f(t)e^{-t/u}dt, \,\;\ q_1, q_2>0,\,\,u\in(q_1,q_2).
\end{eqnarray}
As far as the convergence, $E[f(t)]$ converges in an interval containing $u=0$, provided that (i) $f^{(n)}(0)\rightarrow 0$ as $n\rightarrow \infty$ and (ii) ${lim}_{n\rightarrow \infty}|\frac{f^{(n+1)}(0)}{f^{(n)}(0)}u|<1$. This means the convergence radius $r={lim}_{n\rightarrow \infty}|\frac{f^{(n)}(0)}{f^{(n+1)}(0)}|$ of $E[f(t)]$ depends on the sequence $f^{(n)}(0)$. The superscript $(n)$ of the function is used for the meaning of $n$-th derivative with respect to the concerned variable. \\
The derivative property of Elzaki transform [18-19] is expressed as 
\begin{eqnarray}
E[f^{(n)}(t)]=\frac{T(u)}{u^n}-\sum_{k=0}^{n-1}u^{2-n+k}f^{(k)}(0)\,.
\end{eqnarray} 
Moreover using Eq.(2.1) and Eq.(2.2) few general relations of derivative can be expressed as  
\begin{subequations}
\begin{align}
E[tf^{(n)}(t)]&=u^2\frac{d}{du}E[f^{(n)}(t)]-uE[f^{(n)}(t)]\,,\\
E[t^2f^{(n)}(t)]&=u^4\frac{d^2}{du^2}E[f^{(n)}(t)]\,,\\ 
E[t^3f^{(n)}(t)]&=u^6\frac{d^3}{du^3}E[f^{(n)}(t)]+3u^5\frac{d^2}{du^2}E[f^{(n)}(t)]\,.
\end{align}
\end{subequations}   
The very important fact of Elzaki transform is that, it is dual of Laplace transform. In this context, if $T(u)$ be Elzaki transform of the function $f(t)$ in\\ $\mathcal{A}=\left\{f(t)|\exists M,q_1,q_2>0 \,\,such\,\ that\,\ |f(t)|<M e^{\frac{|t|}{q_j}}, \,\ if\,\ t\in (-1)^j\times[0,\infty)\right\}$ with Laplace transform $F(s)=L[f(t)]$, then $T(u)=uF(\frac{1}{u})$ [20]. This is very useful relation to find the Elzaki transform of any function, if Laplace transform of the given function is known within the domain. To illustrate elaborately, here we have few examples\\
(1) $L[cost]=\frac{s}{s^2+1}$ corresponds to $E[cost]=u\frac{1/u}{[\frac{1}{u^2}+1]}=\frac{u^2}{1+u^2}$ \\
(2) $L[t^n]=\frac{n!}{s^{n+1}}$ corresponds to $E[t^n]=u\frac{n!}{(1/u)^{n+1}}=n!u^{n+2}$ and \\
(3) $L[e^{at}]=\frac{1}{s-a}$ corresponds to $E[e^{at}]=u[\frac{1}{\frac{1}{u}-a}]=\frac{u^2}{1-au}$.\\
The inverse Elzaki transform is given by $f(t)=E^{-1}[T(u)]$. The best way to find the inverse transform is the convolution theorem.\\ According to this, if we have $E[f(t)]=T(u)=\frac{1}{u}g(u)h(u)$ where $g(u)=E[G(t)]$ and $h(u)=E[H(t)]$ then
\begin{eqnarray}
E[(G*H)(t)]=\frac{1}{u}g(u)h(u)\,,
\end{eqnarray}
where 
\begin{eqnarray}
(G*H)(t)=\int_0^{t}G(t-\tau)H(\tau)d\tau\,.
\end{eqnarray}
hence the original function $f(t)=E^{-1}[T(u)]=(G*H)(t)$.\\
This concludes the very short introduction of Elzaki transform which is enough for the present paper. For further one can go through the appendix section. 
\subsection{Formula based approach to deal with $N$-dimensional Schr\"{o}dinger equation}
Schr\"{o}dinger equation is a second order differential equation with non-constant coefficients under the potential term $V(r)$. In this subsection we will use Elzaki transform to develop a formula based approach to deal with multidimensional Schr\"{o}dinger equation for many famous solvable potential models. The term ``solvable" means the solution of the particular potential model provides a closed form     
\begin{eqnarray*}
\,^{(N)}\psi_{n\ell m}\equiv\psi(r, \theta_{1}, \theta_{2}, \ldots, \theta_{N-2},
\phi)=\sum_{n, \ell,
m}C_{n\ell N}R_{n\ell N}(r)Y_{\ell}^{m}(\theta_{1}, \theta_{2},
\ldots, \theta_{N-2}, \phi)\,,
\end{eqnarray*}
where $\,^{(N)}\psi_{n\ell m}$ is the solution of Eq.(3.1) and $R_{n\ell N}(r)$ is the solution of Eq.(3.5). All other symbols have their appropriate meaning, which will be explored in the next section. The radial eigenfunction is the important part of our study specially for radially symmetric potentials. The wavefunction $\psi$ supposed to follow the well behaved boundary condition, namely $\psi(r\rightarrow 0)\rightarrow 0$ and $\psi(r\rightarrow \infty)\rightarrow 0$ for bound states spectrum of chosen potential model, which is one of the prime goal to achieve at least for practical implication of the model.\\
Now it is time to elaborate the step by step methods about how to study Elzaki transform on the above boundary condition for a specific class of potential.\\
{\bf step-I}\\
The boundary condition on $\psi$ is actually associated with the radial part $R_{n\ell N}(r)$ (in short $R(r)$ for convenience). So after inserting the potential into the radial Schr\"{o}dinger equation, a transformation $R(r)=\phi(r)f(r)$ is employed with a  trust that  $R(r\rightarrow 0)\rightarrow 0$ and $R(r\rightarrow \infty)\rightarrow 0$.    
The function $\phi(r)$ controls the asymptotic nature of the radial eigenfunctions. There are few choices for $\phi(r)$ like $r^{-k}|_{k>0}$ or $r^{k}|_{k<0}$. We will take $\phi(r)=r^{-k}|_{k>0}$ for our study. Hence the introduction of  $R(r)=r^{-k}f(r)$ to the radial equation generates a differential equation of $f(r)$ that may contain singular term at the origin. The function $f(r)$ is unknown but surely it is expected to follow the rule $f(r\rightarrow 0)=0$. It is worth to mention here that the parameter $k$ is very crucial for obtaining the overall picture of the concern problem under investigation. $k$ is determined from the differential equation of $\chi(y)|_{y=\eta(r)}\equiv f(r)$ by imposing the condition for "removal of singularity". The new variable $y$ not only helps to interrelate the different potential parameters as well as it guides the differential equation towards the way of easy Elzaki transformation. The ultimate goal of step-I is to obtain a special type of differential equation of the form 
\begin{eqnarray}
y\frac{d^2\chi(y)}{dy^2}+A\frac{d\chi(y)}{dy}+(C-B^2y)\chi(y)=0\,,
\end{eqnarray}
where $A,B(>0),C$ are real free parameters. We will name this equation as MDE (model differential equation). It is to be mentioned here that, non-introduction or introduction of new variable $y$ purely depends on the mathematical situation whether the MDE emerges for $f(r)$ automatically or not. In section 3 we will see that for Coulomb or Mie-type potential introduction of new variable is not necessary whereas it is a must situation for pseudoharmonic, harmonic type potential.  \\
{\bf step-II}\\
Now we have to solve the Eq.(2.6) by Elzaki transform under the initial condition $\chi(0)=0$ as $f(0)=0$. Defining $E[\chi(y)]=T(u)$, Elzaki transform of the differential equation yields 
\begin{eqnarray}
\frac{dT(u)}{T(u)}=\frac{\frac{3-A}{u}-B^2u-C}{1-B^2u^2}du=\frac{\frac{3-A}{u^3}-\frac{C}{u^2}-\frac{B^2}{u}}{\frac{1}{u^2}-B^2}du\,,
\end{eqnarray}
where, following the equation set (2.3), we have used $E[y\frac{d^2\chi(y)}{dy^2}]=\frac{dT(u)}{du}-\frac{3}{u}T(u)$,\\ $E[\frac{d\chi(y)}{dy}]=\frac{T(u)}{u}$ and $E[y\chi(y)]=u^2\frac{dT(u)}{du}-uT(u)$  \\
Taking $\frac{1}{u}=z$, it is easy to find 
\begin{eqnarray}
T\Big(\frac{1}{z}\Big)=K\frac{1}{z}(z+B)^{\frac{1}{2}(A-2-\frac{C}{B})}(z-B)^{\frac{1}{2}(A-2+\frac{C}{B})}\nonumber\\ T(u)=Ku\Big(\frac{1}{u}+B\Big)^{\frac{1}{2}(A-2-\frac{C}{B})}\Big(\frac{1}{u}-B\Big)^{\frac{1}{2}(A-2+\frac{C}{B})}\,,
\end{eqnarray}
where $K$ is the integration constant. Introducing new symbol $p=-\frac{1}{2}(A-2+\frac{C}{B})$ and reinserting $z=\frac{1}{u}$, it is straight forward to reach
\begin{eqnarray}
T(u)=K\frac{1}{u}g(u)h(u)\,,
\end{eqnarray} 
where $g(u)=\frac{u^{p+\frac{C}{B}+1}}{(1+Bu)^{p+\frac{C}{B}}}$ and $h(u)=\frac{u^{p+1}}{(1-Bu)^{p}}$.\\
{\bf step-III}\\
The term $B$ is positive real number. Since $u$ is positive, the second factor $h(u)$ could become negative if $0<\frac{1}{u}<B$, thus its denominator's power must be a negative integer to get single valued eigenfunction. This will also exclude the possibility of getting singularity in the transformation. So we have the quantization condition as $p=-n,(n \in\mathbb Z_+)$. This quantization condition facilitates the energy eigenvalues for investigating potential model. \\
{\bf step-IV}\\
Here from the inverse Elzaki transform via convolution theorem, can take us to the solution in real space. To use the convolution theorem we have
\begin{align}
G(y)&=E^{-1}[g(u)]=\frac{y^{p+\frac{C}{B}-1}e^{-By}}{\Gamma(p+\frac{C}{B})}\,,\\
H(y)&=E^{-1}[h(u)]=\frac{y^{p-1}e^{By}}{\Gamma p}\,.
\end{align}
The original function $\chi(y)$ can be extracted as\\ $\chi(y)=E^{-1}[T(u)]=K(G*H(y)=K\int_{0}^{y}G(y-\tau)H(\tau)d\tau $ which after calculation delivers
\begin{eqnarray}
\chi(y)=K\frac{e^{-By}}{\Gamma(p+\frac{C}{B})\Gamma p}\mathcal{B}\Big(p+\frac{C}{B},p\Big)y^{2p+\frac{C}{B}-1}\,_{1}F_{1}(p,2p+\frac{C}{B},2By)\,,
\end{eqnarray}
where the following formula [21]
\begin{eqnarray}
\int_{0}^x (x-\zeta)^{\sigma-1}\zeta^{\rho-1} e^{\delta\zeta}d\zeta=\mathcal{B}(\sigma,\rho)x^{\sigma+\rho-1}\,_{1}F_{1}(\rho, \sigma+\rho, \delta x)\,.
\end{eqnarray}
has been used. The symbols $\mathcal{B}(\sigma,\rho), \,_{1}F_{1}$ are used for beta function and confluent hypergeometric function respectively. Since $\mathcal{B}(\sigma,\rho)=\frac{\Gamma(\sigma)\Gamma(\rho)}{\Gamma(\sigma+\rho)}$  we have from Eq.(2.12)
\begin{eqnarray}
\chi(y)=K\frac{e^{-By}}{\Gamma(2p+\frac{C}{B})}y^{2p+\frac{C}{B}-1}\,_{1}F_{1}(p,2p+\frac{C}{B},2By)\,.
\end{eqnarray} 
To end this, the solution of radial equation is written as $R(r)=r^{-k}\chi(y)|_{\stackrel{y=\eta(r)}{p=-n}}$. 
\section{A\lowercase{pplication of the method}}
In this section we will use the results of section 2 for the finding of bound states of $N$ dimensional Schr\"{o}dinger equation for various potential models. The $N$-dimensional time independent Schr\"{o}dinger equation for a particle of mass $M$ with orbital angular momentum quantum number
$\ell$ is given by [22]
\begin{eqnarray}
\Bigg[\nabla_N^{2}+\frac{2M}{\hbar^2}\Big(E-V(r)\Big)\Bigg]\,^{(N)}\psi_{n\ell m}(r,\Omega_N)=0\,,
\end{eqnarray}
where $E$ and  $V(r)$ denote the energy eigenvalues and potential. $\Omega_N$ within the argument of $n$-th state eigenfunctions $\,^{(N)}\psi_{n\ell m}$ denotes angular variables
$\theta_1,\theta_2,\theta_3,.....,\theta_{N-2},\varphi$. The Laplacian operator in hyperspherical coordinates is written as
\begin{eqnarray}
\nabla_N^{2}=\frac{1}{r^{N-1}}\frac{\partial}{\partial r}\left(r^{N-1}\frac{\partial}{\partial r}\right)-\frac{\Lambda_{N-1}^{2}}{r^2}\,,
\end{eqnarray}
where
\begin{eqnarray}
\Lambda_{N-1}^{2}=-\Bigg[\sum_{k=1}^{N-2}\frac{1}{sin^2\theta_{k+1}sin^2\theta_{k+2}.....sin^2\theta_{N-2} sin^2\varphi}\times\left(\frac{1}{sin^{k-1}\theta_k}\frac{\partial}{\partial \theta_k}sin^{k-1}\theta_k\frac{\partial}{\partial \theta_k}\right)\nonumber\\+\frac{1}{sin^{N-2}\varphi}\frac{\partial}{\partial\varphi}sin^{N-2}\varphi\frac{\partial}{\partial\varphi}\Bigg]\,.
\end{eqnarray}
$\Lambda_{N-1}^{2}$ is known as the hyperangular momentum operator.\\
We chose the bound state eigenfunctions $\,^{(N)}\psi_{n\ell m}(r,\Omega_N)$ that are vanishing for $r\rightarrow0$ and $r\rightarrow\infty$. Applying the separation variable method by means of the solution
$\,^{(N)}\psi_{n\ell m}(r,\Omega_N)=R_{n\ell N}(r)Y_\ell^{m}(\Omega_N)$,  Eq.(3.1) provides two separated equations
\begin{eqnarray}
\Lambda_{N-1}^{2}Y_\ell^{m}(\Omega_N)=\ell(\ell+N-2)Y_\ell^{m}(\Omega_N)\,,
\end{eqnarray}
where $Y_\ell^{m}(\Omega_N)$ is known as the hyperspherical harmonics and the
hyperradial or in short the ``radial'' equation
\begin{eqnarray}
\left[\frac{d^2}{dr^2}+\frac{N-1}{r}\frac{d}{dr}-\frac{\ell(\ell+N-2)}{r^2}-\frac{2M}{\hbar^2}[V(r)-E]\right]R_{n\ell N}(r)=0\,,
\end{eqnarray}
where $\ell(\ell+N-2)|_{N>1}$ is the separation constant [23] with $\ell=0, 1,2, \ldots$ 
\subsection{Coulomb potential}
The renowned Coulomb potential is written $V(r)=-\frac{Ze^2}{r}$, where $Z, e$ are the atomic number of the matter and electronic charge respectively. $r$ denotes the radial distance of the electron from the center of the atom. Inserting this potential into the Eq.(3.5) and assuming a solution like $R_{n\ell N}(r)=r^{-k_1}f(r)(k_1>0)$ as mentioned in step-I we have
\begin{eqnarray}
r\frac{d^2f}{dr^2}+(N-1-2k_1)\frac{df}{dr}+\Big[\frac{Q_1(k_1,N,\ell)}{r}+\alpha-\beta^2r\Big]f=0\,,
\end{eqnarray}
where $Q_1(k_1,N,\ell)=k_1(k_1+1)-k_1(N-1)-\ell(\ell+N-2)$, $\beta^2=-\frac{2ME}{\hbar^2}$ and $\alpha=\frac{2MZe^2}{\hbar^2}$.
The value of $k_1$ can be obtained from the condition of removal of singularity of the Eq.(3.6) and this yields
\begin{eqnarray}
Q_1(k_1,N,\ell)=0\,.
\end{eqnarray}   
The accepted solution is $k_1(>0)=\ell+N-2$. Hence we have 
\begin{eqnarray}
r\frac{d^2f}{dr^2}-(2\ell+N-3)\frac{df}{dr}+[\alpha-\beta^2r]f=0\,.
\end{eqnarray}
On defining $E[f(r)]=T(u)$ and properly realizing the factors $A,B,C$ of MDE, the solution reads from step-II as
\begin{eqnarray}
T(u)=K_{01}\frac{1}{u}g(u)h(u)\,,
\end{eqnarray}
where $g(u)=\frac{u^{p+\frac{\alpha}{\beta}+1}}{(1+\beta u)^{p+\frac{\alpha}{\beta}}}$, $h(u)=\frac{u^{p+1}}{(1-\beta u)^{p}}$ with
$p=\frac{(2\ell+N-1)\beta-\alpha}{2\beta}$ and $K_{01}$ plays the role of integration constant. The quantization condition of step-III yields
\begin{eqnarray}
p=\frac{(2\ell+N-1)\beta-\alpha}{2\beta}=-n \,\, \mbox{with}\,\ n=0,1,2...
\end{eqnarray}
Eventually, the solutions become from step-IV
\begin{eqnarray}
f(r)=K_{01}\frac{e^{-\beta r}}{\Gamma (2\ell+N-1)}r^{2\ell+N-2}\,_{1}F_{1}(-n, 2\ell+N-1, 2\beta r)\,.
\end{eqnarray}   
The complete radial eigenfunctions $R_{n\ell N}(r)=r^{-k_1}f(r)$ are now
\begin{eqnarray}
R_{n\ell N}(r)=C_{01}r^{\ell} e^{-\beta r}\,_{1}F_{1}(-n, 2\ell+N-1, 2\beta r)\,,
\end{eqnarray}  
where the term $C_{01}=\frac{K_{01}}{\Gamma (2\ell+N-1)}$ is the normalization constant which can be extracted from the condition 
$\int_{0}^{\infty}|R_{n\ell N}(r)|^2 r^{N-1}dr=1$. The energy eigenvalue from Eq.(3.10) comes out as
\begin{eqnarray}
E_{n\ell N}=-\frac{MZ^2e^4}{2\hbar^2}\frac{1}{\Big[n+\ell+\frac{N-1}{2}\Big]^2}\,.
\end{eqnarray}
These results are exactly the same, which were derived in [6].
\subsection{Mie-type potential}
In the study of diatomic molecules Mie-type of potentials arise very frequently. The general form of which is taken as
$V(r)=\frac{a}{r^2}+\frac{b}{r}+c$. This class of potential provides modified Kartzer potential $\Big[V(r)=-D_0(\frac{r-r_0}{r})^2\Big]$ with $a=-D_0r_0^2, b=2D_0r_0$ and $c=-D_0$ and Kratzer-Fues potential $\Big[V(r)=-D_0(\frac{2r_0}{r}-\frac{r_0^2}{r^2})\Big]$ with $a=D_0r_0^2 , b=-2D_0r_0$ and $c=0$. The parameters $D_0$  is the interaction energy between two atoms in a molecular system at equilibrium $r=r_0$. Now following step-I, selecting the solution $R_{n\ell N}(r)=r^{-k_2}f(r), (k_2>0)$ for Eq.(3.5) and imposing the parametric condition
\begin{eqnarray}
Q_2(k_2,N,\ell)=k_2(k_2+1)-k_2(N-1)-\nu(\nu+1)=0\,,
\end{eqnarray}
to remove the singularity in the differential equation, it is not hard to find the MDE  
\begin{eqnarray}
r\frac{d^2f}{dr^2}-(2k_2^{\ell N}-N+1)+(\gamma-\epsilon^2 r)f=0\,.
\end{eqnarray}
Here $k_2^{\ell N}$ is the positive root of Eq.(3.14). All other symbols of Eq.(3.14) and Eq.(3.15) have the following interconnecting relations with the potential parameters
\begin{subequations}
\begin{align}
\nu(\nu+1)&=\ell(\ell+N-2)+\frac{2Ma}{\hbar^2}\,,\\
\gamma&=-\frac{2Mb}{\hbar^2}\,,\\
\epsilon^2&=-\frac{2M(E-c)}{\hbar^2}\,.
\end{align}
\end{subequations}
 If we define $T(u)=E[f(r)]$ and properly realize $A,B,C$ of MDE, then it is easy to reach
\begin{eqnarray}
T(u)=K_{02}\frac{1}{u}g(u)h(u)\,,
\end{eqnarray} 
where $g(u)=\frac{u^{p+\frac{\gamma}{\epsilon}+1}}{(1+\epsilon u)^{p+\frac{\gamma}{\epsilon}}}$, $h(u)=\frac{u^{p+1}}{(1-\epsilon u)^{p}}$ with
$p=\frac{2k_2^{\ell N}-N+3}{2}-\frac{\gamma}{2\epsilon}$ and $K_{02}$ acts the role of integration constant. The quantization condition $(p=-n, n=0,1,2,3\cdots)$ enables to get the energy eigenvalues as
\begin{eqnarray}
E_{n\ell N}=c-\frac{M}{2\hbar^2}\Bigg(\frac{b}{n+k_2^{\ell N}+\frac{3-N}{2}}\Bigg)^2\,.
\end{eqnarray}
The step-IV now provides
\begin{eqnarray}
f(r)=\frac{K_{02}}{\Gamma(2k_2^{\ell N}-N+3)}e^{-\epsilon r}r^{2k_2^{\ell N}-N+2}\,_{1}F_{1}(-n,2k_2^{\ell N}-N+3, 2\epsilon r)\,.
\end{eqnarray} 
The complete radial eigenfunctions are
\begin{eqnarray}
R_{n\ell N}(r)=r^{-k_2^{\ell N}}f(r)=C_{02}e^{-\epsilon r}r^{k_2^{\ell N}-N+2}\,_{1}F_{1}(-n,2k_2^{\ell N}-N+3, 2\epsilon r)\,,
\end{eqnarray}
where $C_{02}=\frac{K_{02}}{\Gamma(2k_2^{\ell N}-N+3)}$ is the normalization constant which is possible to evaluate from the condition $\int_{0}^{\infty}|R_{n\ell N}(r)|^2 r^{N-1}dr=1$. The results are well matched with the previous works[11].
\subsection{Harmonic potential}
Simple harmonic oscillator potential is one of the first exactly solvable famous potential. In terms of mass $M$ and circular frequency $\omega$ of a particle the potential is written $V(r)=\frac{1}{2}M\omega^2r^2$, where $r$ defines the displacement of the particle from its equilibrium position. As in step-I, inserting this potential and the solution form $R_{n\ell N}(r)=r^{-k_3}f(r),(k_3>0)$ into Eq.(3.5) we have
\begin{eqnarray}
\frac{d^2f}{dr^2}+\frac{N-1-2k_3}{r}\frac{df}{dr}+\Big[\frac{Q_3(k_3,N,\ell)}{r^2}+\beta_0^2-\alpha_0^2 r^2\Big]f=0\,,
\end{eqnarray}  
where $Q_3(k_3,N,\ell)=k_3(k_3+1)-k_3(N-1)-\ell(\ell+N-2)$, $\beta_0^2=\frac{2ME}{\hbar^2}$ and $\alpha_0=\frac{M\omega}{\hbar}$. This equation is not the same as the MDE. So to achieve the form here it is useful to introduce a new variable $y=r^2$. This immediately gives
\begin{eqnarray}
y\frac{d^2\chi}{dy^2}+\Big(\frac{N}{2}-k_3\Big)\chi+\Big(\frac{1}{4}\beta_0^2-\frac{1}{4}\alpha_0^2 y\Big)\chi=0\,,
\end{eqnarray} 
where $\chi(y)|_{y=r^2}=f(r)$ and the singularity removal condition 
\begin{eqnarray}
Q_3(k_3,N,\ell)=0\,,
\end{eqnarray}
has been used as the guide line of step-I. The accepted solution of $k_3(>0)$ is $\ell+N-2$. On realizing $T(u)=E[\chi(y)]$ and properly realize $A,B,C$ of MDE, then it is easy to reach
\begin{eqnarray}
T(u)=K_{03}\frac{1}{u}g(u)h(u)\,,
\end{eqnarray} 
where $g(u)=\frac{u^{p+\frac{\beta_0^2}{2\alpha_0}+1}}{(1+\frac{\alpha_0}{2}u)^{p+\frac{\beta_0^2}{2\alpha_0}}}$, $h(u)=\frac{u^{p+1}}{(1-\frac{\alpha_0}{2} u)^{p}}$ with
$p=\frac{1}{2}[\frac{N}{2}+\ell-\frac{\beta_0^2}{2\alpha_0}]$ and $K_{03}$ acts the role of integration constant. Once again the the quantization condition $(p=-n, n=0,1,2,3\cdots)$ facilitates the energy eigenvalue as
\begin{eqnarray}
E_{n\ell N}=\hbar \omega\Big(2n+\ell+\frac{N}{2}\Big)\,.
\end{eqnarray}
The expression for $\chi(y)$ emerges from step-IV as
\begin{eqnarray}
\chi(y)=K_{03}\frac{e^{-\frac{\alpha_0}{2}y}}{\Gamma(2p+\frac{\beta_0^2}{2\alpha_0})}y^{2p+\frac{\beta_0^2}{2\alpha_0}-1}\,_{1}F_{1}(p, 2p+\frac{\beta_0^2}{2\alpha_0}, \alpha_0 y)\,.
\end{eqnarray}
The actual radial eigenfunctions are now
\begin{eqnarray}
R_{n\ell N}(r)=r^{-k_3}\chi(y)|_{\stackrel{y=r^2}{p=-n}}=C_{03}e^{-\frac{\alpha_0}{2}r^2}r^{\ell}\,_{1}F_{1}(-n, \frac{N}{2}+\ell, \alpha_0 r^2)\,,
\end{eqnarray}
where we have used $C_{03}=\frac{K_{03}}{\Gamma(\frac{N}{2}+\ell)}$ is the normalization constant to be evaluated from the condition $\int_{0}^{\infty}|R_{n\ell N}(r)|^2 r^{N-1}dr=1$. The results are those what obtained in the ref.[7].
\subsection{Pseudoharmonic potential}
The pseudoharmonic potential is used to describe the roto-vibrational states of diatomic molecules
as well as for nuclear rotations and vibrations. The potential is written in the standard form
\begin{eqnarray}
V(r)=a_1r^2+\frac{a_2}{r^2}+a_3\,,
\end{eqnarray}  
where $a_1=\frac{D_e}{r_e^2}, a_2=D_er_e^2$ and $a_3=-2D_e$ will take the Eq.(3.28) into its original form
$V(r)=D_e\Big(\frac{r}{r_e}-\frac{r_e}{r}\Big)^2$[24]. Here $D_e=\frac{1}{8}K_er_e^2$ is the dissociation energy with the force constant $K_e$ and $r_e$ is the equilibrium constant. Inserting the potential expressed by Eq.(3.28) into Eq.(3.5) and adopting the following abbreviations
\begin{subequations}
\begin{align}
\nu(\nu+1)&=\ell(\ell+N-2)+\frac{2M}{\hbar^2}a_1\,,\\
\mu^2&=\frac{2M}{\hbar^2}a_1\,,\\
\epsilon_0^2&=\frac{2M(E-a_3)}{\hbar^2}\,,
\end{align}
\end{subequations}
we have 
\begin{eqnarray}
\frac{d^2R_{n\ell N}(r)}{dr^2}+\frac{N-1}{r}\frac{dR_{n\ell N}(r)}{dr}+\Big[\epsilon_0^2-\frac{\nu(\nu+1)}{r^2}-\mu^2r^2\Big]R_{n\ell N}(r)=0\,.
\end{eqnarray}
Taking the expected solution $R_{n\ell N}(r)=r^{-k_4}f(r), (k_4>0)$ as previous and changing the variable $y=r^2$ and realizing $f(r)\equiv \chi(y)|_{y=r^2}$ we have
\begin{eqnarray}
4y\frac{d^2\chi}{dy^2•}+2(N-2k_4)\frac{d\chi}{dy}+\Big[\frac{Q_4(k_4,N,\ell)}{y}-\mu^2y+\epsilon_0^2\Big]\chi=0\,,
\end{eqnarray}
where $Q_4(k_4,N,\ell)=k_4(k_4+1)-k_4(N-1)-\nu(\nu+1)$.
Now to free from the singular term we impose a parametric condition
\begin{eqnarray}
Q_4(k_4,N,\ell)=0\,,
\end{eqnarray}
which also allow us to get
\begin{eqnarray}
y\frac{d^2\chi}{dy^2}-\Big(k_4^{\ell N}-\frac{N}{2}\Big)\frac{d\chi}{dy}+(\frac{\epsilon_0^2}{4}-\frac{\mu^2}{4}y)\chi=0\,,
\end{eqnarray} 
where $k_4^{\ell N}$ is the positive root of the Eq.(3.32). Here from realizing $T(u)=E[\chi(y)]$ we have the solution in transformed space (just we did in step-II)
\begin{eqnarray}
T(u)=K_{04}\frac{1}{u}g(u)h(u)\,,
\end{eqnarray} 
where $g(u)=\frac{u^{p+\frac{\epsilon_0^2}{2\mu}+1}}{(1+\frac{\mu}{2} u)^{p+\frac{\epsilon_0^2}{2\mu}}}$, $h(u)=\frac{u^{p+1}}{(1-\frac{\mu}{2} u)^{p}}$ with
$p=\frac{k_4^{\ell N}-\frac{N}{2}+2}{2}-\frac{\epsilon_0^2}{4\mu}$ and $K_{04}$ acts the role of integration constant.
The quantization condition $p=-n$ of step-III delivers 
\begin{eqnarray}
E_{n\ell N}=\frac{\hbar^2}{2M}\,\epsilon_0^2+a_3=a_3+\sqrt{\frac{8\hbar^2a_{1}}{M}\,}\left[n+\frac{1}{2}
+\frac{1}{4}\sqrt{(N+2\ell-2)^2+\frac{8Ma_{2}}{\hbar^2}\,}\right]\,,
\end{eqnarray}
where a little algebra has been done with the help of equation set (3.29). Finally moving to step-IV it is easy to find 
\begin{eqnarray}
\chi(y)=K_{04}\frac{e^{-\frac{\mu }{2}y}}{\Gamma(2p+\frac{\epsilon_0^2}{2\mu})}y^{2p+\frac{\epsilon_0^2}{2\mu}-1}\,_{1}F_{1}(p,2p+\frac{\epsilon_0^2}{2\mu},\mu y)\,.
\end{eqnarray}
The radial eigenfunctions are
\begin{eqnarray}
R_{n\ell N}(r)=r^{-k_4^{\ell N}}\chi(y)|_{\stackrel{y=r^2}{p=-n}}=C_{04}e^{-\frac{\mu}{2}r^2}r^{k_4^{\ell N}-N+2}\,_{1}F_{1}(-n,k_4^{\ell N}-\frac{N}{2}+2,\mu r^2) \,,
\end{eqnarray}
where $C_{04}=\frac{K_{04}}{\Gamma(k_4^{\ell N}-\frac{N}{2}+2)}$ is the normalization constant which is easy to derive from the condition $\int_{0}^{\infty}|R_{n\ell N}(r)|^2 r^{N-1}dr=1$.  
These all results are consistent with the ref.[12]
\section{c\lowercase{onclusion}}
In this paper Elzaki transform is first time used to find the solution of Schr\"{o}dinger equation for different potential models, specially for $N$-dimension. We have shown a general scheme, $R(r)=\phi(r)f(r)=r^{-k}f(r)|_{k>0}$, is very suitable for applying Elzaki transform to specific class of potential. The solutions have been obtained in closed form. This study emerges the Elzaki transform as a new tool for solving Schr\"{o}dinger equation as well as related problems.\\
The study directly does not provide the one dimensional spectrum of the investigated potentials as the entire mathematics is restricted for $N>1$. One dimensional Schr\"{o}dinger equation should be studied separately and we checked that the general scheme is equally applicable there. Apart form the investigated potentials there are more potential models like the generalized Morse potential, the inverse square potential $\frac{A}{x^2}$, the central-force model of deuteron $V(x)=-Ae^{-x}$, the radial equation of the two-dimensional Schr\"{o}dinger equation for free particle follow the same path of this paper. For more complicated class of potential, the universal scheme will also be effective but mathematical realization and cleverness is required for the first function $\phi$. Here we have few examples in tabular form   
\begin{table}[htpb]
\begin{center}
\caption{Complicated potentials and the universal Elzaki transformation scheme}
\vspace{5mm}
\begin{tabular}{c|c}\hline
Potential $V(x)$ & Transform Scheme $\psi(\xi)\equiv\phi(\xi)f(\xi)$ \\ \hline
P\"{o}schl-Teller potential: $Acsc^2 x, 0\leq x\leq\pi$ & $(1-\xi)^{k}f(\xi), \xi=cos(2x)$ \\ \hline 
Hulth\'{e}n potential: $\frac{A}{2}[1-coth(x/2)], 0\leq x\leq \infty$ & $ (1+\xi)^{-k_1}(1-\xi)^{k_2}f(\xi), \xi=coth(x/2)$ \\ \hline
step potential: $A/2[1+tanh(x/2)], -\infty\leq x\leq \infty$ & $(1+\xi)^{ik_1}(1-\xi)^{k_2}f(x), \xi=tanh(x/2)$ \\ \hline
Modified P\"{o}schl-Teller potential: $-A sech^2(\alpha x), -\infty\leq x\leq \infty$& $ (1-\xi^2)^{k}f(\xi), \xi=tanh(\alpha x)$ \\ \hline
\end{tabular}
\end{center}
\end{table}\\
Certainly for this tabular potentials one has to deal with differential equation containing term like $(1-\xi^2)f^{''}(\xi)$ and $\xi f^{'}(\xi)$. Elzaki transform of those equations become very easy in transformed space. It is to inform  here that above potentials have been already studied in ref.[25] by Laplace transform. So as a dual of Laplace transform surely Elzaki transform will also solve the equations successfully. \\
In general mathematics, compared to Laplace and Sumudu transforms Elzaki transform is more suitable to solve the ordinary differential equation with variable coefficients, but this does not demand that Elzaki transform always provides a general solution of a problem. For instant seeking for a solution to the following initial value problem $(y^{''}+xy^{'}(x)-y=0\,\,\, y(0)=0, y^{'}(0)=1)$ one gets $\frac{dT(u)}{du}+(\frac{1}{u^3}-\frac{3}{u})T(u)=1$ where $E[y(x)]=T(u)$. Now $T(u)=u^3$ and $u^3+cu^3e^{\frac{1}{2u^2}}$ both satisfy the transformed differential equation but a non zero $c$ makes the inverse transform complicated. If one choose $c=0$ or the first choice then solution becomes $y(x)=E^{-1}[u^3]=x$. \\
In case of quantum mechanical problems, combination of universal transform scheme as well as the boundary condition makes the Elzaki transform smooth and guides the solution into a general closed form automatically. This subject, namely use of Elzaki transform in quantum problems, has just stepped in with this paper. We are looking forward to the future research for more development of it.    
\section*{R\lowercase{eferences}}

\newpage
\small
\renewcommand{\theequation}{A-\arabic{equation}}
\setcounter{equation}{0}  
\section*{Appendix}
\begin{table}[htpb]
\begin{center}
\caption{Elzaki transform of some functions}
\vspace{10mm}
\begin{tabular}{|c|c|}\hline
$f(t)$ & $T(u)=E\left\{f(t)\right\}$   \\ \hline
$t^n$ & $n!u^{n+2}$ \\ \hline 
$\frac{t^{a-1}}{\Gamma(a)}, a>0$ & $ u^{a+1}$ \\ \hline
$e^{at}$ & $\frac{u^2}{1-au}$ \\ \hline
$\frac{t^{n-1}e^{at}}{(n-1)!}, n=1,2,3.\cdots$ & $\frac{u^{n+1}}{(1-au)^n}$  \\ \hline
$sin(at)$&$ \frac{au^3}{1+a^2u^2}$\\ \hline
$cos(at)$&$\frac{u^2}{1+a^2u^2}$\\ \hline
$sinh(at)$&$\frac{au^3}{1-a^2u^2}$\\ \hline
$cosh(at)$&$\frac{au^2}{1-a^2u^2}$\\ \hline
$e^{at}sin(bt)$&$\frac{bu^3}{(1-au)^2+b^2u^2}$\\ \hline
$e^{at}cos(bt)$&$ \frac{(1-au)u^2}{(1-au)^2+b^2u^2}$ \\ \hline
$tsin(at)$&$\frac{2au^4}{1+a^2u^2}$\\ \hline
$H(t-a)$&$ u^2e^{-\frac{a}{u}}$ \\ \hline
$\delta(t-a)$&$ ue^{-\frac{a}{u}}$\\ \hline
$J_0(at)$&$ \frac{u^2}{\sqrt{1+a^2u^2}}$\\ \hline
$ \frac{(t-a)^{n-1}}{\Gamma(n)}H(t-a)$& $u^{n+1}e^{-\frac{a}{u}}$\\ \hline
\end{tabular}
\end{center}
\end{table}
\begin{center}
{\bf A\\The solution of the Bessel equation of zeroth order $xy^{''}(x)+y^{'}(x)+a^2xy(x)=0, y(0)=1$}
\end{center}
Defining $E\left\{y(x)\right\}=T(u)$, Elzaki transform on both side of the equation yields 
\begin{eqnarray}
u^2\frac{d}{du}\Big[\frac{T(u)}{u^2}-1-uk_0\Big]-u\Big[\frac{T(u)}{u^2}-1-uk_0\Big]+\frac{T(u)}{u}-u+a^2\Big[u^2\frac{dT(u)}{du}-uT(u)\Big]=0\,,
\end{eqnarray}
where $k_0=f^{'}(0)$. Arranging the terms it is not hard to reach
\begin{eqnarray}
\frac{dT(u)}{T(u)}=\frac{1}{1+a^2u^2}\Big[\frac{2}{u}+a^2u\Big]du\,.
\end{eqnarray}
Integrating both side 
\begin{eqnarray}
T(u)=C\frac{u^2}{\sqrt{1+a^2u^2}}\,,
\end{eqnarray}
where $C$ contains the integration constant within it.
The solution of $y(x)$ can be found by inverse Elzaki transform and it is 
\begin{eqnarray}
y(x)=E^{-1}\left\{T(u)\right\}=CJ_0(ax)\,.
\end{eqnarray}
The two-dimensional radial Schr\"{o}dinger equation for the free particle is a Bessel equation. The result given by Eq.(A-4) may be useful for zeroth order solution of it.
\small
\renewcommand{\theequation}{B-\arabic{equation}}
\setcounter{equation}{0}
\begin{center}
{\bf B\\ Solution of $\frac{d^2y(x)}{dx^2}+\omega^2 y(x)=0$ with the condition $y(0)=a , y^{'}(0)=b$}
\end{center}
Defining $E\left\{y(x)\right\}=T(u)$ and after transformation of the differential equation we have
\begin{eqnarray}
\frac{T(u)}{u^2}-y(0)-uy^{'}(0)+\omega^2 T(u)=0\,,
\end{eqnarray}
which gives 
\begin{eqnarray}
T(u)=\frac{au^2+bu^3}{1+\omega^2 u^2}=\frac{au^2}{1+\omega^2 u^2}+\frac{bu^3}{1+\omega^2 u^2}\,.
\end{eqnarray}
In real space the solution is 
\begin{eqnarray}
y(x)=E^{-1}\left\{T(u)\right\}=acos(\omega x)+\frac{b}{\omega}sin(\omega x)\,.
\end{eqnarray}
The solution is well known to literature. It is as important in acoustics as in quantum mechanical
elementary text book problem, where the potential (infinite square well) is defined 
\begin{eqnarray}
V(x)=\begin{cases} 0,& \text{if }  0\leq x\leq a\\
\infty .& \text{ otherwise} 
\end{cases}
\end{eqnarray}
The Schr\"{o}dinger equation simply comes out as $\frac{d^2\psi}{dx^2}+k^2\psi=0,$ where energy $E>0$ and $k^2=\frac{2ME}{\hbar^2}$.
\small
\renewcommand{\theequation}{C-\arabic{equation}}
\setcounter{equation}{0}
\begin{center}
{\bf C\\  Shifting theorem: }
If $E\left\{f(x)\right\}=T(u)$ then $E\left\{e^{-ax}f(x)\right\}=(1+au)T\Big(\frac{u}{1+au}\Big)$, where $a$ is real
\end{center}
By definition
\begin{eqnarray}
E\left\{e^{-ax}f(x)\right\}=u\int_0^{\infty}e^{-\frac{1+au}{u}x}f(x)dx=\frac{p}{1-ap}\int_0^{\infty}e^{-\frac{x}{p}}f(x)dx=\frac{1}{1-ap}T(p)\,,
\end{eqnarray}
where $p=\frac{u}{1+au}$ or $u=\frac{p}{1-ap}$. This finally delivers
\begin{eqnarray}
E\left\{e^{-ax}f(x)\right\}=(1+au)T\Big(\frac{u}{1+au}\Big)\,.
\end{eqnarray}

\begin{thebibliography}{99}
\bibitem{ref1} Debnath L and Bhatta D 2006 \textit{Integral transform and their application}(Boca Raton: CRC Press) 
\bibitem{ref2} Arfken G B and Weber H J 1995 \textit{Mathematical Methods for Physicists} (San Diego: Academic Press)
\bibitem{ref3} Schr\"{o}dinger E 1926 Ann. Physik {\bf79} 361
\bibitem{ref4} Swainson R A and Drake G W F 1991 J. Phys.A {\bf 24} 79
\bibitem{ref5} Englefield M J 1974 J. Math. Anal. Appl {\bf 48} 270
\bibitem{ref6} Chen G 2004 Z Naturforsch {\bf 59a} 875
\bibitem{ref7} Gang C 2005 Chin.Phy {\bf14} 1075 
\bibitem{ref8} Arda A and Sever R 2012  J. Math. Chem {\bf50} 971
\bibitem{ref9} Miraboutalebi S and Rajaei L 2014 J.Math.Chem {\bf 52} 1119
\bibitem{ref10} Pimentel D R M and de Castro A S 2013 Euro. J. Phys {\bf34} 199
\bibitem{ref11} Das T 2015 J. Math. Chem.{\bf53} 618  
\bibitem{ref12} Das T and Arda A 2015 Adv. High Energy Phys. {\bf 2015} 137038
\bibitem{ref13} Watugala G K 1993 Int. J. Math. Edu. Sci. Tech {\bf24} 35
\bibitem{ref14} Asiru M A 2002 Int. J. Math. Edu. Sci. Tech {\bf33} 441
\bibitem{ref15} Elzaki T M 2011 Global .J. pure and Appl. Math {\bf7} 57
\bibitem{ref16} Elzaki T M, Elzaki S M and Elnour E A 2012 Global. J. Math. Sci: Theo and Prac {\bf 4} 1
\bibitem{ref17} Kim H 2013 Int. J. pure and Appl. Math {\bf87} 261
\bibitem{ref18} Elzaki T M, Elzaki S M and Elnour E A 2012 Global. J. Math. Sci: Theo and Prac {\bf 4} 15 
\bibitem{ref19} Elzaki T M, Elzaki S M 2011 Advances in Theo. Appl. Math {\bf6} 13
\bibitem{ref20} Elzaki T M, Elzaki S M 2011 Advances in Theo. Appl .Math {\bf6} 1
\bibitem{ref21} Gradshteyn I S and Ryzhik I M 2007 \textit{Table of integrals, Series, and Products} (Academic Press: New York)
\bibitem{ref22} Chatterjee A 1990 Phys.Rep. {\bf186} 249
\bibitem{ref23} Shimakura N 1992 \textit{Partial Differential Operator of Elliptic Type} (American Math Society: Providence)
\bibitem{ref24} Goldman I and Krivchenkov V 1960 \textit{ Problems in Quantum Mechanics} (Pergamon Press: London)
\bibitem{ref25} Tsaur G Y and Wang J 2014 Eur. J. Phys {\bf35} 015006
\end{thebibliography}
\end{document}